\documentclass[english,aps,preprint]{revtex4}
\usepackage[T1]{fontenc}
\usepackage[latin9]{inputenc}
\usepackage{color}
\usepackage{amsmath}
\usepackage{amssymb}

\makeatletter

\@ifundefined{textcolor}{}
{%
 \definecolor{BLACK}{gray}{0}
 \definecolor{WHITE}{gray}{1}
 \definecolor{RED}{rgb}{1,0,0}
 \definecolor{GREEN}{rgb}{0,1,0}
 \definecolor{BLUE}{rgb}{0,0,1}
 \definecolor{CYAN}{cmyk}{1,0,0,0}
 \definecolor{MAGENTA}{cmyk}{0,1,0,0}
 \definecolor{YELLOW}{cmyk}{0,0,1,0}
}

\makeatother

\usepackage{babel}
\begin{document}
\title{Supplemental material for ``Theoretical bound of the efficiency of\textcolor{red}{{}
}\textcolor{black}{learning}''}
\author{Shanhe Su$,^{1,\;*}$ Ousi Pan}
\email{These authors contributed equally to this work.}
\address{Department of Physics, Xiamen University, Xiamen, 361005, People's
	Republic of China}
\author{Shihao Xia}
\address{Department of Physics, Xiamen University, Xiamen, 361005, People's
	Republic of China}
\author{Jincan Chen}
\email{jcchen@xmu.edu.cn}
\address{Department of Physics, Xiamen University, Xiamen, 361005, People's
	Republic of China}
\author{Chikako Uchiyama}
\address{Faculty of Engineering, University of Yamanashi, Kofu, Yamanashi 400-8511,
	Japan and National Institute of Informatics, Chiyoda, Tokyo 101-8430,
	Japan}
\begin{abstract}
The supplementary information is organized as follows. In Sec. I,
on the basis of the Markovian master equation and the time derivative
of Shannon's entropy, the Clausius inequality of the bipartite system
is derived. In Sec. II, by following the calculation for the bipartite
system, the Clausius inequality of a subsystem will be revealed based
on the coarse-grained dynamics. The rate of entropy production of
the subsystem will be splited into three components, including the
entropy-production rate, time derivative of the Shannon entropy, and
\textcolor{black}{rate of information learned.} In Sec. III, the lower
limit of the rate of entropy production of the subsystem is evaluated,
and is transformed to obtain the universal upper bound for the efficiency
of learning. The basic thermodynamics quantities for
the DQD system and the cellular network are presented in Sec. IV.
\end{abstract}
\maketitle

\section{Thermodynamics of a bipartite system }

The Clausius inequality of the bipartite system $Z$ can be identified
by utilizing the Markovian master equation {[}Eq. (1) in the main
text{]}. First, the time derivative of the Shannon entropy $S=-\sum_{z}p(z)\ln p(z)$
of system $Z$ \cite{Nielsen2010,Landi2021} yields
\begin{align}
\dot{S} & =-\sum_{z}\dot{p}(z)\ln p(z)\nonumber \\
 & =-\sum_{z,z^{\prime},v}\left[W_{zz^{\prime}}^{v}p(z^{\prime})-W_{z^{\prime}z}^{v}p(z)\right]\ln p(z)\nonumber \\
 & =-\sum_{z,z^{\prime},v}J_{zz^{\prime}}^{v}\ln p(z)\nonumber \\
 & =\frac{1}{2}\sum_{z,z^{\prime},v}J_{zz^{\prime}}^{v}\ln\frac{p(z^{\prime})}{p(z)}\nonumber \\
 & =\frac{1}{2}\sum_{z,z^{\prime},v}J_{zz^{\prime}}^{v}\ln\left[\frac{p(z^{\prime})W_{zz^{\prime}}^{v}}{p(z)W_{z^{\prime}z}^{v}}\right]-\frac{1}{2}\sum_{z,z^{\prime},v}J_{zz^{\prime}}^{v}\ln\left(\frac{W_{zz^{\prime}}^{v}}{W_{z^{\prime}z}^{v}}\right)\nonumber \\
 & =\dot{\sigma}-\dot{S_{r}}\text{.}\label{eq:sn}
\end{align}
For a Markov process with the local detailed-balance condition,
\begin{equation}
\dot{S_{r}}=\frac{1}{2}\sum_{z,z^{\prime},v}J_{zz^{\prime}}^{v}\ln\left(\frac{W_{zz^{\prime}}^{v}}{W_{z^{\prime}z}^{v}}\right)
\end{equation}
is the rate of entropy change of the environment, which is associated
with the energy flow to the environment \cite{Schaller2014,Shiraishi2019}.
As $\dot{\sigma}$ is the sum of the rates of entropy change in the
system and the environment, i.e.,

\begin{equation}
\dot{\sigma}=\dot{S}+\dot{S_{r}}\text{,}\label{eq:epr}
\end{equation}
one figures out that $\dot{\sigma}$ is the rate of entropy production.
By combining Eqs. (\ref{eq:sn}) and (\ref{eq:epr}), $\dot{\sigma}$
has the following equivalent form \cite{Schnakenberg1976}

\begin{equation}
\dot{\sigma}=\frac{1}{2}\sum_{z,z^{\prime},v}J_{zz^{\prime}}^{v}\ln\left[\frac{p(z^{\prime})W_{zz^{\prime}}^{v}}{p(z)W_{z^{\prime}z}^{v}}\right]\geq0\text{.}\label{eq:tentropy}
\end{equation}
It is always non-negative because of the form $\left(x-y\right)\ln\left(\frac{x}{y}\right)\geqslant0$
\cite{Landi2021}, meaning that the evolution described by Eq. (1)
in the main text satisfies the second law of thermodynamics.

\section{Thermodynamics of the subsystems}

In this section, the Clausius inequality of the subsystems will be
revealed based on the coarse-grained dynamics {[}Eq. (3) in the main
text{]}. First of all, \textcolor{black}{the rate of information learned}
is calculated by the time derivative of the mutual information $I$.
The quantity $I$ representing the correlation between subsystems
$X$ and $Y$ is expressed as
\begin{equation}
I=\sum_{x,y}p(z)\ln\frac{p(z)}{p(x)p(y)}\geq0\text{,}
\end{equation}
where $p(x)=\sum_{y}p(z)=\sum_{y}p(x,y)$ and $p(y)=\sum_{x}p(z)=\sum_{x}p(x,y)$
are the marginal probabilities of the states of subsystems $X$ and
$Y$, respectively. It follows from Jensen\textquoteright s inequality
that $I$ is always nonnegative \cite{Cover1999}. The time derivative
of $I$ is divided as
\begin{equation}
\dot{I}=\dot{I}^{X}+\dot{I}^{Y}
\end{equation}
with
\begin{equation}
\dot{I}^{X}=\frac{1}{2}\sum_{x,x^{\prime},y,v}J_{xx^{\prime}\mid y}^{v}\ln\frac{p\left(y|x\right)}{p\left(y|x^{\prime}\right)}\label{eq:IX}
\end{equation}
and

\begin{equation}
\dot{I}^{Y}=\frac{1}{2}\sum_{x,y,y^{\prime},v}J_{yy^{\prime}\mid x}^{v}\ln\frac{p\left(x|y\right)}{p\left(x|y^{\prime}\right)}.
\end{equation}
Note that

\begin{equation}
J_{xx^{\prime}\mid y}^{v}=W_{xx^{\prime}\mid y}^{v}p(x^{\prime},y)-W_{x^{\prime}x\mid y}^{v}p(x,y)
\end{equation}
is the current due to the transition of $X$ from state $x^{\prime}$
to state $x$ provided that subsystem $Y$ is at state $y$. Similarly,
\begin{equation}
J_{yy^{\prime}\mid x}^{v}=W_{yy^{\prime}\mid x}^{v}p(x,y^{\prime})-W_{y^{\prime}y\mid x}^{v}p(x,y)
\end{equation}
is the current due to the transition of $Y$ from state $y^{\prime}$
to state $y$ when subsystem $X$ is at state $x$. As explained in
the main text, $\dot{I}^{X}$ and $\dot{I}^{Y}$ describe the rates
of learning between $X$ and $Y$.

In order to relate the Clausius inequality of a subsystem to the \textcolor{black}{rate
of information learned}, we start from the evolution of the Shannon
entropy $S^{X}=-\sum_{x}p(x)\ln p(x)$ of subsystem $X$. Differentiating
$S^{X}$ with respect to time and inserting the master equation of
coarse-grained states of $X$ {[}Eq. (3) in the main text{]} lead
to
\begin{align}
\dot{S}^{X} & =-\sum_{x}\dot{p}(x)\ln p(x)\nonumber \\
 & =-\sum_{x,x^{\prime},v}\left[V_{xx^{\prime}}^{v}p\left(x^{\prime}\right)-V_{x^{\prime}x}^{v}p\left(x\right)\right]\ln p_{x}\nonumber \\
 & =-\sum_{x,x^{\prime},y,v}J_{xx^{\prime}\mid y}^{v}\ln p(x)\nonumber \\
 & =\frac{1}{2}\sum_{x,x^{\prime},y,v}J_{xx^{\prime}\mid y}^{v}\ln\frac{p(x^{\prime})}{p(x)}\nonumber \\
 & =\frac{1}{2}\sum_{x,x^{\prime},y,v}J_{xx^{\prime}\mid y}^{v}\ln\frac{p(x^{\prime},y)W_{xx^{\prime}\mid y}^{v}}{p(x,y)W_{x^{\prime}x\mid y}^{v}}-\frac{1}{2}\sum_{x,x^{\prime},y,v}J_{xx^{\prime}\mid y}^{v}\ln\frac{W_{xx^{\prime}\mid y}^{v}}{W_{x^{\prime}x\mid y}^{v}}\nonumber \\
 & +\frac{1}{2}\sum_{x,x^{\prime},y,v}J_{xx^{\prime}\mid y}^{v}\ln\frac{p(x^{\prime})p(x,y)}{p(x)p(x^{\prime},y)}\nonumber \\
 & =\frac{1}{2}\sum_{x,x^{\prime},y,v}J_{xx^{\prime}\mid y}^{v}\ln\frac{p(x^{\prime},y)W_{xx^{\prime}\mid y}^{v}}{p(x,y)W_{x^{\prime}x\mid y}^{v}}-\frac{1}{2}\sum_{x,x^{\prime},y,v}J_{xx^{\prime}\mid y}^{v}\ln\frac{W_{xx^{\prime}\mid y}^{v}}{W_{x^{\prime}x\mid y}^{v}}\nonumber \\
 & +\frac{1}{2}\sum_{x,x^{\prime},y,v}J_{xx^{\prime}\mid y}^{v}\ln\frac{p\left(y|x\right)}{p\left(y|x^{\prime}\right)}.\label{eq:SX}
\end{align}
The conditional probabilities $p\left(y|x^{\prime}\right)=p\left(x^{\prime},y\right)/p\left(x^{\prime}\right)$
and $p\left(y|x\right)=p\left(x,y\right)/p(x)$ have been inserted
to obtain the last equality.

Under the condition of local detailed balance,
\begin{equation}
\dot{S_{r}}^{X}=\frac{1}{2}\sum_{x,x^{\prime},y,v}J_{xx^{\prime}\mid y}^{v}\ln\frac{W_{xx^{\prime}\mid y}^{v}}{W_{x^{\prime}x\mid y}^{v}}\label{eq:snX}
\end{equation}
is the rate of entropy change in the environment associated with the
energy flow from subsystem $X$. By following Eq. (\ref{eq:tentropy}),
the entropy-production rate of subsystem $X$ is identified as

\begin{equation}
\dot{\sigma}^{X}=\frac{1}{2}\sum_{x,x^{\prime},y,v}J_{xx^{\prime}\mid y}^{v}\ln\frac{p\left(x^{\prime},y\right)W_{xx^{\prime}\mid y}^{v}}{p\left(x,y\right)W_{x^{\prime}x\mid y}^{v}}\geq0.\label{eq:epx-1}
\end{equation}
The positivity of Eq. (\ref{eq:epx-1}) is deduced by recognizing
that it has the form $\left(x-y\right)\ln\left(\frac{x}{y}\right)\geqslant0$
as well. Using Eqs. (\ref{eq:IX}) and (\ref{eq:SX})-(\ref{eq:epx-1}),
one can split $\dot{\sigma}^{X}$ into three components, i.e.,

\begin{equation}
\dot{\sigma}^{X}=\dot{S}^{X}+\dot{S_{r}}^{X}-\dot{I}^{X}\geq0.\label{eq:epr-1}
\end{equation}

For the same reason, the entropy-production rate of subsystem $Y$

\begin{equation}
\dot{\sigma}^{Y}=\dot{S}^{Y}+\dot{S_{r}}^{Y}-\dot{I}^{Y}\geq0\text{,}
\end{equation}
where the time derivative of the Shannon entropy of subsystem $Y$
\begin{equation}
\dot{S}^{Y}=-\sum_{y}\dot{p}(y)\ln p(y),
\end{equation}
and the rate of entropy change in the environment associated with the
energy flow from subsystem $Y$
\begin{equation}
\dot{S_{r}}^{Y}=\frac{1}{2}\sum_{x,y,y^{\prime},v}J_{yy^{\prime}\mid x}^{v}\ln\frac{W_{yy^{\prime}\mid x}^{v}}{W_{y^{\prime}y\mid x}^{v}}\text{.}
\end{equation}

\section{the effectiveness of the informational learning }

In this section, we derive the limit for the efficiency of learning.
First, the upper bound of $\left|\dot{S_{r}}^{X}\right|$ is evaluated
as follows
\begin{align}
\left|\dot{S_{r}}^{X}\right|= & \left|\frac{1}{2}\sum_{x,x^{\prime},y,v}J_{xx^{\prime}\mid y}^{v}\ln\frac{W_{xx^{\prime}\mid y}^{v}}{W_{x^{\prime}x\mid y}^{v}}\right|\nonumber \\
= & \left|\frac{1}{2}\sum_{x,x^{\prime},y,v}\ln\frac{W_{xx^{\prime}\mid y}^{v}}{W_{x^{\prime}x\mid y}^{v}}\sqrt{W_{xx^{\prime}\mid y}^{v}p\left(x^{\prime},y\right)+W_{x^{\prime}x\mid y}^{v}p(x,y)}\frac{J_{xx^{\prime}\mid y}^{v}}{\sqrt{W_{xx^{\prime}\mid y}^{v}p\left(x^{\prime},y\right)+W_{x^{\prime}x\mid y}^{v}p(x,y)}}\right|\nonumber \\
\overset{(a)}{\leq} & \frac{1}{2}\sqrt{\sum_{x,x^{\prime},y,v}\left(\ln\frac{W_{xx^{\prime}\mid y}^{v}}{W_{x^{\prime}x\mid y}^{v}}\right)^{2}\left[W_{xx^{\prime}\mid y}^{v}p\left(x^{\prime},y\right)+W_{x^{\prime}x\mid y}^{v}p(x,y)\right]}\sqrt{\sum_{x,x^{\prime},y,v}\frac{\left(J_{xx^{\prime}\mid y}^{v}\right)^{2}}{W_{xx^{\prime}\mid y}^{v}p\left(x^{\prime},y\right)+W_{x^{\prime}x\mid y}^{v}p(x,y)}}\nonumber \\
\overset{(b)}{\leq} & \frac{1}{2}\sqrt{\sum_{x,x^{\prime},y,v}\left(\ln\frac{W_{xx^{\prime}\mid y}^{v}}{W_{x^{\prime}x\mid y}^{v}}\right)^{2}\left[W_{xx^{\prime}\mid y}^{v}p\left(x^{\prime},y\right)+W_{x^{\prime}x\mid y}^{v}p(x,y)\right]}\sqrt{\frac{1}{2}\sum_{x,x^{\prime},y,v}J_{xx^{\prime}\mid y}^{v}\ln\frac{W_{xx^{\prime}\mid y}^{v}p\left(x^{\prime},y\right)}{W_{x^{\prime}x\mid y}^{v}p(x,y)}}\nonumber \\
= & \frac{1}{2}\sqrt{\sum_{x,x^{\prime},y,v}\left(\ln\frac{W_{xx^{\prime}\mid y}^{v}}{W_{x^{\prime}x\mid y}^{v}}\right)^{2}\left[W_{xx^{\prime}\mid y}^{v}p\left(x^{\prime},y\right)+W_{x^{\prime}x\mid y}^{v}p(x,y)\right]}\sqrt{\dot{\sigma}^{X}}\nonumber \\
= & \sqrt{\frac{1}{2}\sum_{x,x^{\prime},y,v}\left(\ln\frac{W_{xx^{\prime}\mid y}^{v}}{W_{x^{\prime}x\mid y}^{v}}\right)^{2}W_{xx^{\prime}\mid y}^{v}p\left(x^{\prime},y\right)}\sqrt{\dot{\sigma}^{X}}\nonumber \\
= & \sqrt{\mathcal{A}^{X}\dot{\sigma}^{X}},\label{eq:ins}
\end{align}
where the Cauchy-Schwartz inequality \cite{Shiraishi2018,Funo2019}
and the inequality $\frac{(x-y)^{2}}{x+y}\leq\frac{x-y}{2}\log\frac{x}{y}$
for non-negative $x$ and $y$ have been applied in $\left(a\right)$
and $\text{\ensuremath{\left(b\right)}}$, respectively. With the
help of the inequality Eq. (\ref{eq:ins}), it is natural to obtain
the lower limit on the entropy-production rate of subsystem $X$ {[}Eq.
(10) in the main text{]}. At the steady state, using Eq. (8) in the
main text and squaring Eq. (\ref{eq:ins}) on both sides, we obtain
the following relation
\begin{equation}
\left(\dot{S_{r}}^{X}\right)^{2}\leq\mathcal{A}^{X}\left(\dot{S_{r}}^{X}-\dot{I}^{X}\right).\label{eq:srxq}
\end{equation}

By considering that $\dot{S_{r}}^{X}\geq\dot{I}^{X}\geq0$, Eq. (\ref{eq:srxq})
obviously satisfies
\begin{equation}
\dot{S_{r}}^{X}\leq\mathcal{A}^{X}\left(1-\frac{\dot{I}^{X}}{\dot{S_{r}}^{X}}\right).\label{eq:Srxin}
\end{equation}
At the last step, by introducing the definition of the efficiency
of learning, Eq. (\ref{eq:Srxin}) is transformed to reach the universal
upper bound for the efficiency of learning, i.e.,
\begin{equation}
\eta^{X}=\frac{\dot{I}^{X}}{\dot{S_{r}}^{X}}\leq1-\frac{\dot{S_{r}}^{X}}{\mathcal{A}^{X}}\text{.}
\end{equation}

\section{The basic thermodynamic quantities for the DQD system
and the cellular network}

For the DQD system, the entropy flow associated with
subsystem $X$
\begin{equation}
\dot{S_{r}}^{X}=-\beta(\mu_{XR}-\mu_{XL})\left(J_{x_{0}x_{1}\mid y_{0}}^{XR}+J_{x_{0}x_{1}\mid y_{1}}^{XR}\right)-\beta U\left(J_{y_{1}y_{0}\mid x_{0}}^{YL}+J_{y_{1}y_{0}\mid x_{0}}^{YR}\right),
\end{equation}
the rate of information learned

\begin{equation}
\dot{I}^{X}=\left(J_{y_{1}y_{0}\mid x_{0}}^{YL}+J_{y_{1}y_{0}\mid x_{0}}^{YR}\right)\ln\frac{p\left(x_{0},y_{1}\right)p\left(x_{1},y_{0}\right)}{p\left(x_{0},y_{0}\right)p\left(x_{1},y_{1}\right)},
\end{equation}
and the coefficient
\begin{align}
\mathcal{A}^{X} & =\frac{1}{2}\left(\varepsilon_{X}-\mu_{XL}\right)^{2}\left[W_{10\mid0}^{XL}p\left(0,0\right)+W_{01\mid0}^{XL}p\left(1,0\right)\right]\nonumber \\
 & +\frac{1}{2}\left(\varepsilon_{X}-\mu_{XR}\right)^{2}\left[W_{10\mid0}^{XR}p\left(0,0\right)+W_{01\mid0}^{XR}p\left(1,0\right)\right]\nonumber \\
 & +\frac{1}{2}\left(\varepsilon_{X}+U-\mu_{XL}\right)^{2}\left[W_{10\mid1}^{XL}p\left(0,1\right)+W_{01\mid1}^{XL}p\left(1,1\right)\right]\nonumber \\
 & +\frac{1}{2}\left(\varepsilon_{X}+U-\mu_{XR}\right)^{2}\left[W_{10\mid1}^{XR}p\left(0,1\right)+W_{01\mid1}^{XR}p\left(1,1\right)\right].
\end{align}

For the cellular network, the entropy flow to the
reservoir associated with the protein (subsystem $X$)
\begin{equation}
\dot{S_{r}}^{X}=J\Delta\mu,\label{eq:csr}
\end{equation}
the rate of information learned

\begin{equation}
\dot{I}^{X}=J\ln\frac{p\left(x_{1},y_{1}\right)p\left(x_{0},y_{0}\right)}{p\left(x_{0},y_{1}\right)p\left(x_{1},y_{0}\right)},\label{eq:cI}
\end{equation}
and the coefficient
\begin{align}
\mathcal{A}^{X} & =\frac{1}{2}\left(\ln\frac{\kappa_{+}}{\kappa_{-}}\right)^{2}\left[\kappa_{-}p\left(x_{1},y_{1}\right)+\kappa_{+}p\left(x_{0},y_{1}\right)\right]\nonumber \\
 & +\frac{1}{2}\left(\ln\frac{\omega_{+}}{\omega_{-}}\right)^{2}\left[\omega_{+}p\left(x_{1},y_{0}\right)+\omega_{-}p\left(x_{0},y_{0}\right)\right].
\end{align}

\end{document}